\documentstyle[prl,psfig,aps,multicol]{revtex}
\psfigurepath{/month/saif/Part2/latex/}

\begin{document}
\draft

\title{The Fermi accelerator in atom optics}

\author{Farhan Saif\cite{sa}\,, Iwo Bialynicki-Birula\cite{bia}\,, 
        Mauro Fortunato\cite{maur} and 
        Wolfgang P. Schleich}
\address{Abteilung f\"ur Quantenphysik, Universit\"at Ulm,
          Albert-Einstein-Allee 11, D-89069 Ulm, Germany}
\maketitle
\begin{abstract}
We study the classical and quantum dynamics of a Fermi
accelerator realized by an atom bouncing off a modulated atomic mirror.
We find that in a window of the modulation amplitude
dynamical localization occurs in both position and momentum. 
A recent experiment [A. Steane, P. Szriftgiser,
                 P. Desbiolles, and J. Dalibard, Phys. Rev. Lett. {\bf 74}, 4972
(1995)] shows that this system can be implemented experimentally.
\end{abstract}
\begin{multicols}{2}

Almost fifty years ago Enrico Fermi~\cite{kn:fermi} suggested that 
it is collisions with moving magnetic fields that accelerate 
cosmic rays. Since then many models associated with the
name of Fermi accelerator describing 
particles colliding with moving walls 
have been investigated theoretically 
\cite{kn:lieb,kn:chen,kn:flat1,kn:benv,kn:oliv}. 
The recent experiment~\cite{kn:sten} on atoms bouncing off a modulated
atomic mirror constitutes an atom optics \cite{kn:mlyn} version of the 
Fermi accelerator.  
In the present paper, we argue
that this system serves as a testing ground for 
classical and quantum chaos~\cite{kn:haak}.
We show that there exists an
experimentally accessible parameter regime in which such
experiments can bring out many features of the Fermi accelerator.

Three properties make this model rather special~\cite{kn:schlautman} in
the field of chaos: 
(i) It is one of the very few bound systems with no continuum
\cite{kn:benv1,kn:moham},
(ii) the phenomenon of dynamical localization occurs in 
the position {\it and} momentum variables\cite{kn:bard},
(iii) dynamical localization
arises only in a window of the modulation depth;
the onset of classical chaos sets the lower boundary,
whereas the effective dimensionless Planck constant $k^{\hspace{-2.1mm}-}$ 
determines the upper boundary. 
Therefore, we can tune the system continuously from a 
regime of no chaos, through one with classical chaos but 
dynamical localization, to one with classical chaos and
no dynamical localization. The experiments performed so far
have focused on the latter regime. However, we show that they can
easily be extended into the localization window.  

We now consider a cloud of 
laser-cooled atoms stored in a magneto-optical trap.
When we switch off the trap, the atoms move along 
the $\tilde z$-direction~\cite{kn:tildas} under the influence
of gravity and bounce off an atomic mirror~\cite{kn:amin}. 
The latter results from a laser field incident
on a glass prism under an angle of total internal
reflection. This creates an
evanescent wave whose intensity $I(\tilde z)=I_0\exp(- 2k \tilde z)$
decays over a distance $k^{-1}$ outside of the prism. 
A sinusoidal modulation \cite{kn:sten} of the intensity 
with amplitude $\epsilon$ and
frequency $\omega$ changes the intensity $I(\tilde z)$
to $I(\tilde z,\tilde t)=
I_0\exp(- 2k \tilde z)(1+\epsilon\sin\omega\tilde t)$.
In our calculations we have used a slightly different form  
\begin{equation}
I(\tilde z,\tilde t) =
I_0\exp(- 2k\tilde z +\epsilon\sin\omega\tilde t)
\end{equation}
of modulation which corresponds to an oscillation of the mirror and is 
our exponential model of the Fermi accelerator.
When $\epsilon$ is not too large, the results do not 
depend significantly on the form of the modulation. 

In order to avoid problems associated with spontaneous emission we 
consider a large detuning between the laser light field and
the atomic transition frequency. This ensures that the atom 
rarely leaves the ground state.
The dynamics of the center-of-mass motion of the atom in the ground state
follows then from the Hamiltonian
\begin{equation}
H=\frac{\tilde{p}^2}{2m}+mg\tilde{z}
+\frac{\hbar\Omega_{eff}}{4} e^{-2k\tilde{z}+
\epsilon\sin\omega\tilde t}.
\label{ham}
\end{equation}
Here $\tilde p$ is the momentum of the atom of mass $m$
along the $\tilde{z}$-axis, and $g$ 
denotes the gravitational acceleration. 

We introduce the dimensionless position and momentum
coordinates $z\equiv\tilde{z}\omega^2/g$ and $p\equiv\tilde{p}\omega/(mg)$ 
and time $t\equiv\omega\tilde t$ together 
with the dimensionless intensity $V_0\equiv\hbar\omega^2\Omega_{eff}/(4mg^2)$,
steepness $\kappa\equiv 2kg/\omega^2$ and the modulation depth 
$\lambda\equiv\omega^2\epsilon/(2kg)$ of the evanescent wave.
In these variables, the classical dynamics follows from the
Hamilton equations of motion, 
\begin{eqnarray}
\dot z &=& p\nonumber,\\
\dot p &=& -1 + \kappa V_0 \exp\left[-\kappa (z -\lambda\sin t)\right].
\label{eq:hamilt2}
\end{eqnarray}

The corresponding Schr\"odinger equation for the atoms in 
the ground state reads
\begin{equation}
ik^{\hspace{-2.1mm}-}\frac{\partial\psi}{\partial t}=\left[\frac{p^2}{2} + z
+ V_0 \exp\left[-\kappa (z -\lambda\sin t)\right]\right]\psi,
\label{dham}
\end{equation}
where $k^{\hspace{-2.1mm}-}\equiv\hbar\omega^3/(mg^2)$ denotes 
the dimensionless Planck constant~\cite{kn:kbar}. 
 
In this paper, we use Eqs. 
(\ref{eq:hamilt2}) and (\ref{dham}) to determine 
the classical and quantum mechanical
position and momentum distributions of cold atoms 
bouncing under the influence of gravity off an 
oscillating mirror. In all our calculations we start at $t=0$ from an
ensemble of atoms with an average momentum zero and an average position
$z=20$ above the mirror. We find dynamical 
localization in the quantum case. However, 
localization occurs only over a certain range of the modulation
depth $\lambda$. In order to understand this result and to 
find the lower and upper boundaries $\lambda_l$ and $\lambda_u$ of this
regime, we approximate the exponential potential by
a hard wall, that is an infinitely steep and infinitely high wall
and borrow some results\cite{kn:benv,kn:oliv,kn:bren} for such a
time dependent triangular potential well. 

We can approximate the dynamics of this system by a 
map \cite{kn:lieb} connecting two consecutive bounces of the atom:
The momentum $p$ of the atom and the
phase $\theta$ of the wall before a bounce determine 
the momentum $\bar{p}$ and phase
$\bar{\theta}$ just before the next bounce 
through the standard Chirikov-Taylor map
\begin{eqnarray}
\bar p &=& p + K\cos\theta\nonumber\\
\bar\theta&=&\theta+\bar p\,\,\,\,\,\,\,\,{\rm mod(2\pi)}
\label{eq:map}
\end{eqnarray}
with $K=4\lambda$.

Chirikov \cite{kn:chir1} has shown that when the chaos 
parameter $K$ becomes larger than the
critical value $K_{cr}=0.9716...$, the classical 
system undergoes a global diffusion. 
This implies for our driven triangular potential well that diffusive dynamics 
sets in for $\lambda>\lambda_l=K_{cr}/4\approx 0.24$. Below this value
we have isolated resonances, whereas above, 
the resonances overlap and we have islands embedded in a 
stochastic sea. 

In the corresponding quantum mechanical system,
the quasi-energy spectrum of the Floquet operator changes from
a point spectrum to an almost continuum~\cite{kn:oliv} when 
$\lambda>\lambda_u\equiv\sqrt{k^{\hspace{-2.1mm}-}}/2$.
Here quantum diffusion destroys localization.
The conditions of classical and quantum diffusion, together, define the
window 
\begin{equation}
0.24<\lambda<\frac{\sqrt{k^{\hspace{-2.1mm}-}} }{2}
\label{win}
\end{equation}
in the modulation strength, where we can find dynamical localization.

\begin{figure}
\psfig{figure=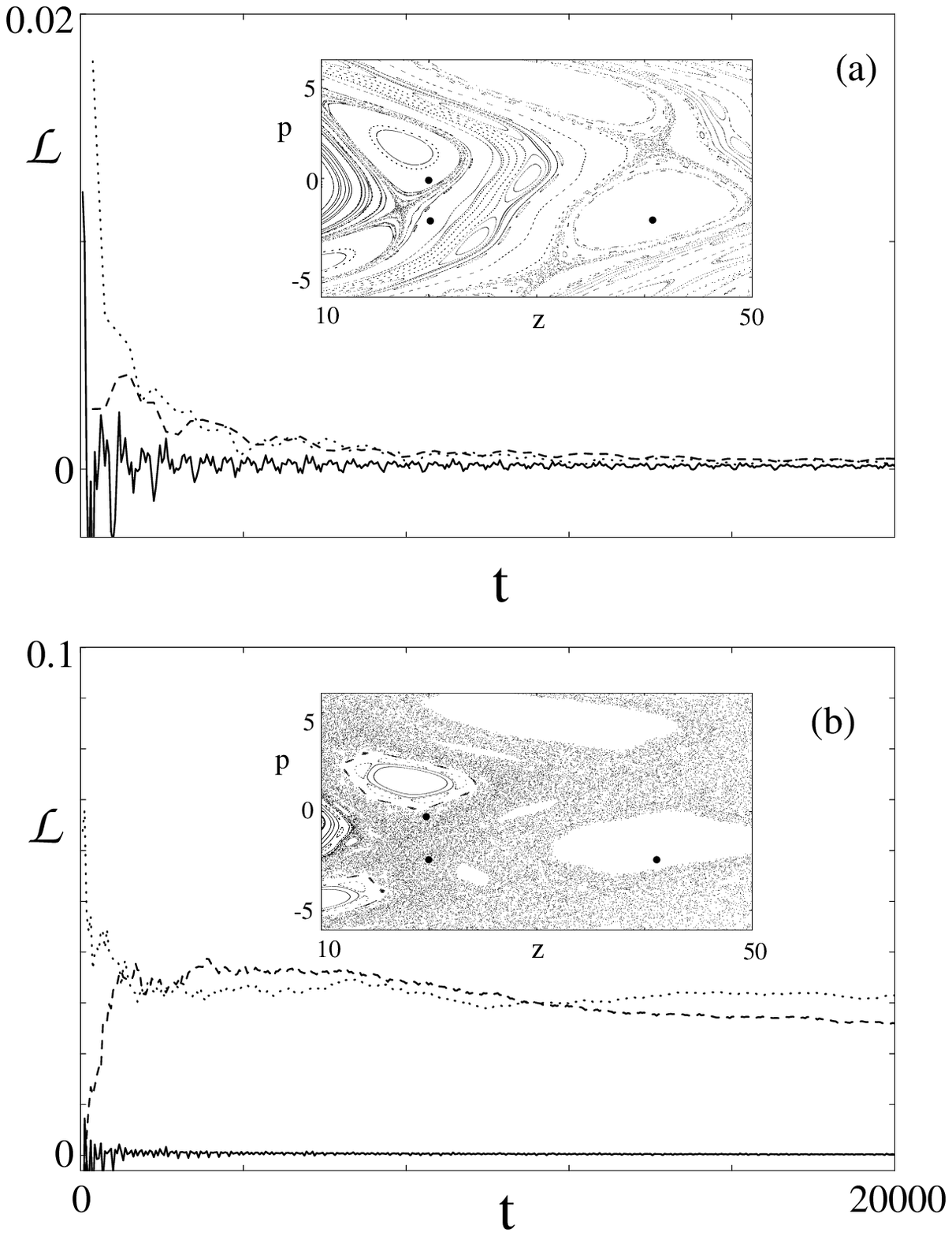,width=3.4in}
\vspace{0.3truecm}
\caption{Lyapunov exponent $\mathcal{L}$ of an atomic 
Fermi accelerator below (a) and above (b) the lower boundary
$\lambda_l=0.24$ of the localization window. We use three initial 
conditions $(z_0,p_0)=\{(20,0);(20,-2);(40,-2)\}$
represented by thick dots in the Poincar\'e sections.
In (a) we have chosen the modulation
depth $\lambda=0.2$ and all initial conditions lie inside 
isolated resonances. Consequently for all three initial conditions
the Lyapunov exponent approaches zero. 
In (b) we have $\lambda=0.5 >\lambda_l$.
Here, the phase space point (40,-2)
still sits in an island whereas the points (20,0) and (20,-2) lie 
in the stochastic sea. As a result the Lyapunov exponent for the first 
initial condition converges to zero whereas for the
other two it is positive. Here and in all other figures we have chosen
for the height of the potential $V_0=60$ and for its steepness $\kappa=0.5$. }
\label{fg:ljap}
\end{figure}

This result obtained for the triangular well 
determines an {\it approximate}
range of $\lambda$ in which we can observe dynamical localization
for the exponential well.
Since the lower boundary $\lambda_l$ is set by classical dynamics
we can find $\lambda_l$ by evaluating
Lyapunov exponents. For a modulation amplitude $\lambda<0.24$, 
the Lyapunov exponent converges to zero in a vast range of initial 
conditions except in small regions near separatrices
as shown in Fig.~\ref{fg:ljap}(a).
However, for larger modulations 
diffusion occurs and the Lyapunov exponent becomes 
positive in a vast range of initial conditions 
as shown in Fig.~\ref{fg:ljap}(b).

\begin{figure}
\psfig{figure=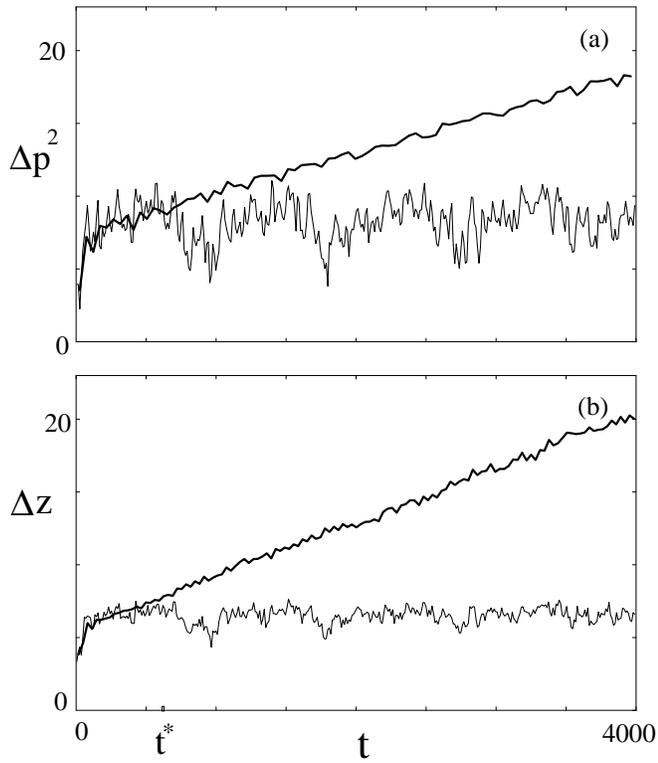,width=3.4in}
\vspace{0.6truecm}
\caption{Comparison between quantum (thin lines) and classical 
(thick lines) values of the (a)
{\it square} of the width $\Delta p$ of 
the momentum distribution and (b) width of the position distribution
$\Delta z$. In the quantum case we integrate the 
Schr\"odinger equation, Eq.~(\ref{dham}), subjected to the initial 
condition of a Gaussian minimum uncertainty wavepacket located at the 
phase space point (20,0) with a width $\Delta z=2$ in position and
the corresponding width 
$\Delta p=k^{\hspace{-2.1mm}-}/(2\Delta z)$ in momentum. 
In the classical case we propagate an ensemble of particles 
distributed according to the same Gaussian distribution  using the classical 
Hamilton equations, Eq.~(\ref{eq:hamilt2}).
The number of particles in the classical simulation is 2000. 
The height of the exponential potential is $V_0=60$, 
its steepness $\kappa=0.5$ and the modulation strength is $\lambda=0.5$.
In the quantum mechanical case the effective Planck's constant
is taken as $k^{\hspace{-2.1mm}-}=4$.
}
\label{fg:wpx}
\end{figure}

Above the lower boundary, the classical system undergoes
diffusion in both position and momentum space. 
This diffusion 
manifests itself in a linear growth of the {\it square} 
$\Delta p^2$ of the width 
$\Delta p\equiv (\langle p^2\rangle-\langle p\rangle^2)^{1/2}$
of the classical momentum distribution and of the 
width $\Delta z\equiv (\langle z^2\rangle-\langle z\rangle^2)^{1/2}$ 
of the classical position distribution with time as shown
in Fig.~\ref{fg:wpx}.  The linear growth of $\Delta p^2$ 
with time, $\Delta p^2\sim Dt$, 
also follows from the Chirikov-Taylor 
map \cite{kn:benv} of Eq.~(\ref{eq:map}).
However, the value of the diffusion constant obtained for our choice
of parameters is much smaller than the value $D=K^2/2$
predicted by the Chirikov-Taylor map.

We note that the
numerical result of Fig.~\ref{fg:wpx} also suggests a linear time 
dependence of the width $\Delta z$ of the position distribution.
In addition we see that the values of $\Delta p^2$ and
$\Delta z$ are approximately equal. These facts can be
explained by assuming that the distribution of positions
and momenta is governed by the classical Boltzmann distribution 
\begin{equation}
P_{cl}(z,p)=(2\pi)^{-1/2}\eta^{-3/2}\exp[ -(p^2/2+ V(z))/\eta]\;,
\label{eq:dist}
\end{equation}
where
\begin{equation}
V(z)=z+ V_0 \exp(-\kappa z).
\end{equation}
\noindent
Here, the quantity $\eta$ depends on time and plays the role of an
effective temperature. This conjecture is supported
by the fact that the Boltzmann statistics holds in case
of diffusive dynamics~\cite{kn:bolt}.

The calculation of the $\Delta z$ and $\Delta p^2$ can
be performed analytically in the simplest case of the
triangular potential well and it gives the equality 
\begin{equation} 
\Delta z=\eta =\Delta p^2\;.
\label{eq:equ}
\end{equation}
We have checked that for an exponential barrier this equality
is also approximately true. 

\begin{figure}
\psfig{figure=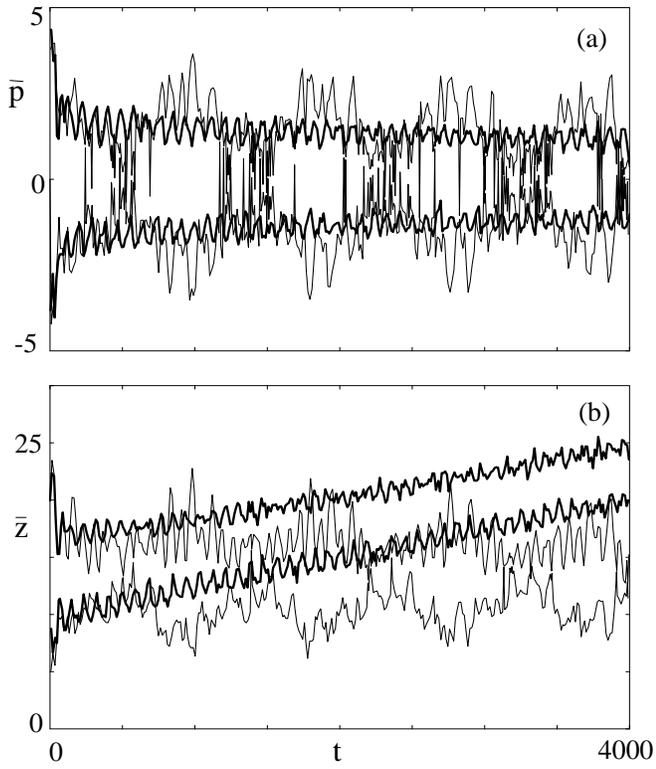,width=3.4in}
\vspace{0.6truecm}
\caption{Comparison between the 
classical (thick lines) and quantum 
(thin lines) average momentum (a) and average position (b) 
as functions of time. For the sake of presentation,
we show the envelope of the corresponding functions. 
We note the modulation of the quantum mechanical 
envelopes which is absent in the classical curves.
This is a manifestation of revivals%
~\protect\cite{kn:clem} 
in a driven quantum system~\protect\cite{kn:saif}.
The calculations were performed for the same
set of parameters as in Fig.~\protect\ref{fg:wpx}.}
\label{fg:dwpx}
\end{figure}

We also show in Fig.~\ref{fg:dwpx}
the average momentum $\bar p$ and average position $\bar z$. 
Classically, the average momentum oscillates around zero, which
corresponds to our initial average momentum, in agreement with the
result obtained from Chirikov map. On the other hand, the 
average position displays a linear rise with time, in
accordance with the relation calculated with the Boltzmann
distribution.  

\begin{figure}
\psfig{figure=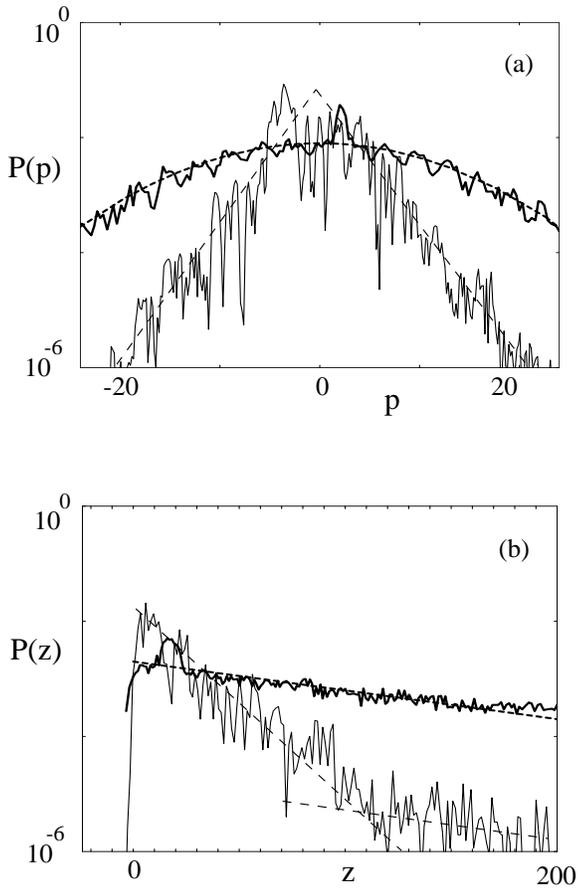,width=3in}
\vspace{0.3truecm}
\caption{Comparison between the quantum mechanical (thin lines) 
and classical (thick lines) momentum (a) and position (b) distributions 
on a logarithmic scale for an atomic Fermi accelerator. 
The momentum distribution in the quantum mechanical case
exhibits an exponential localization, whereas the corresponding 
classical distribution is Gaussian.
The classical 
as well as the quantum mechanical position distributions are both 
of exponential form. However in the 
classical case this form results from the linear potential
in the Boltzmann distribution. The peak around $z=20$
is due to the fact that a considerable part of
our initial ensemble lies inside a stable island. In
the quantum case we find two exponentials:
the flat one is a remnant of the classical Boltzmann distribution and
the steeper one represents
dynamical localization. 
The dashed lines indicate linear and quadratic fits which
correspond to exponential and Gaussian distributions, respectively.
Here we have chosen
$\lambda=0.8$ and the number of particles in the
classical simulation is 10000. The integration time is
$t=2650$ and all the other parameters are the same as
in Fig~\ref{fg:wpx}.
}
\label{fg:dpx}
\end{figure}

The classical 
position and momentum distributions shown in Fig.~\ref{fg:dpx}
follow then the exponential barometric formula 
$P_{cl}(z)=\eta^{-1}\exp(-z/\eta)$
and the Gaussian distribution
$P_{cl}(p)=(2\pi\eta)^{-1/2}\exp[-p^2/(2\eta)]$
predicted from Eq.~(\ref{eq:dist}). 

With the help of these distributions 
we can easily establish the relation $\Delta z=\eta =\Delta p^2$. 
Since we have the diffusion law
$\Delta p^2\sim Dt$, Eq.~(\ref{eq:equ}) predicts that
$\Delta z\sim Dt$ in agreement with the
numerical results of Fig.~\ref{fg:wpx}.

In contrast to this classical diffusion the corresponding 
quantum mechanical quantities saturate after
an initial rise that is of classical nature.
This difference between classical and quantum dynamics manifests
itself after the quantum break time $t^*$. 
We estimate \cite{kn:haak} this time as $t^*\sim D{k^{\hspace{-2.1mm}-}}^{-2}$ 
and the corresponding saturation value of $\Delta p^2$ follows from
$\Delta p^2\sim Dt^*\sim D^2{k^{\hspace{-2.1mm}-}}^{-2}$.

We also note characteristic oscillations in the
quantum mechanical widths $\Delta p^2$ and $\Delta z$.
These oscillations, absent in the classical curves,
are a generalization to driven systems \cite{kn:saif} of the 
revival phenomena \cite{kn:clem}.

The quantum distributions in position and momentum
shown in Fig.~\ref{fg:dpx} are completely different from their
classical counterparts. Indeed the quantum 
mechanical momentum distribution is exponential rather
than Gaussian. Moreover, the quantum mechanical 
position distribution contains two exponentials:
The steep one corresponds to dynamical localization
in position whereas the flat one has the same
steepness as the classical barometric formula.
This separation of the quantum distribution 
into a quantum and a classical part also occurs 
in other bound problems\cite{kn:moham}.

A qualitative comparison between the quantum triangular well and the kicked 
rotator model \cite{kn:haak} yields the localization length as 
$l\sim D {k^{\hspace{-2.1mm}-}}^{-2}\sim 8\lambda^2/{k^{\hspace{-2.1mm}-}}^2$. 
In our numerical investigations we clearly see this qualitative dependence 
on $k^{\hspace{-2.1mm}-}$ and $\lambda$.

\begin{figure}
\psfig{figure=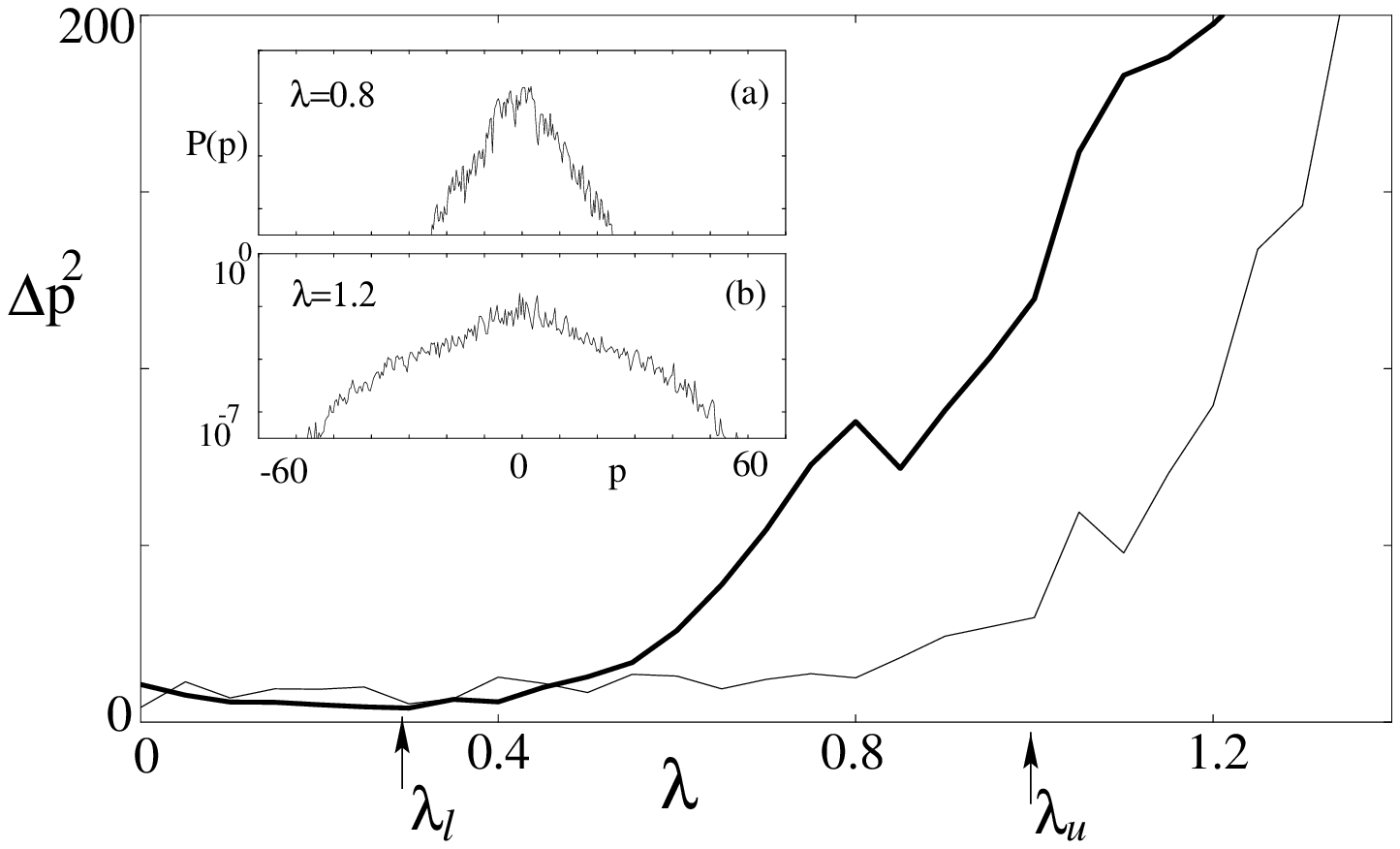,width=3.8in}
\vspace{0.3truecm}
\caption{Localization window defined by the onset of classical
and quantum diffusion at $\lambda_l=0.24$ and
$\lambda_u=\sqrt{k^{\hspace{-2.1mm}-}}/2=1$, respectively. 
Squares of the width of the classical
(thick line) and the quantum mechanical (thin line) momentum 
distributions in their dependence on $\lambda$. 
The two curves start to separate at $\lambda=\lambda_l$. 
A transition from a localized to a delocalized 
quantum mechanical momentum distribution occurs at $\lambda_u$.
For $\lambda=0.8$ which for $k^{\hspace{-2.1mm}-}=4$ 
lies well within the window we find exponential
localization [inset (a)]. In contrast for $\lambda=1.2$ which 
lies outside of the window we find a broad Gaussian 
distribution indicating delocalization [inset (b)]. 
All the other parameters are as in Fig.~\ref{fg:wpx} and
$t=3200$.}
\label{fg:mplocal}
\end{figure}

The initial minimum uncertainty wavepacket delocalizes 
when the modulation depth exceeds the upper boundary
$\lambda_u$ of the localization window.
In Fig.~\ref{fg:mplocal} we show the widths of the
classical (thick line) and the quantum mechanical (thin line)
momentum distribution as a function of the modulation amplitude 
$\lambda$. We find exponential 
localization within the window, as shown in the inset (a). 
However, above the upper boundary $\lambda_u$, quantum diffusion
sets in resulting in the Gaussian distribution 
shown in the inset (b). 

We now make contact with 
the recent experiment\cite{kn:sten} on the phase modulation
of de Broglie waves. In this experiment cesium atoms bounce off a mirror 
with $k^{-1}=0.19~\mu$m. The mirror is modulated with
an amplitude $\epsilon=0.82$ and various frequencies of the order 
$\omega\approx2\pi\times 900$~kHz. 
With the mass $m=2.21\times 10^{-25}$kg of Cs atoms, 
the gravitational
acceleration $g=9.81$~m/s$^2$, and $\hbar=6.673\times 10^{-34}/2\pi$~Js,
we find the dimensionless Planck constant
$k^{\hspace{-2.1mm}-}\approx 9\times 10^8$
and the modulation depth $\lambda\approx 2.5\times 10^5$. 
We emphasize that this value is larger
than $\lambda_u=\sqrt{k^{\hspace{-2.1mm}-}}/2\sim 1.5\times 10^4$. Therefore
this experiment lies outside of the localization window.

However, a modulation frequency of $\omega=2\pi\times 1.477$~kHz,
the decay length $k^{-1}=0.455~\mu$m
and the effective Rabi frequency $\Omega_{eff}=2\pi\times 88.8$~kHz
lead to $k^{\hspace{-2.1mm}-}=4$, $\kappa=0.5$ 
and $V_0=60$ which are the values used 
throughout our paper. We have chosen this value of $V_0$ to guarantee
that the atoms will not hit the surface of the mirror
which is situated at $z=0$. Of course some atoms
have enough energy to break through the barrier but their
number is negligible. The lower and upper boundary
$\lambda_l=0.24$ and $\lambda_u=1$ then translate 
into an intensity modulation of 
$\epsilon_l=0.12$ and $\epsilon_u=0.5$, respectively.

With our parameters we have observed 
dynamical localization within $100$ bounces. 
The quantum break time $t^*\sim 75$ms corresponds to $50$ bounces.
Since Ref.\cite{kn:ovch} reports more than hundred bounces, 
this effect should be observable. 

Note that the initial condition in \cite{kn:sten} was 
$\tilde z_0=3.3$~mm which corresponds to $z_0=2.9\times 10^4$ in dimensionless
coordinates. In contrast, in our numerical calculations we have chosen 
$z_0=20$ which corresponds to $\tilde z=2.27~\mu$m. 
However, a recent experiment\cite{kn:ovch} shows that even this is possible.
We are aware of the fact that the practical implementation
of our proposed experiment may not be trivial. However,
in spite of the experimental challenge---due to the external
modulation of the mirror---we are confident that an
experiment along the above lines can be performed.

We conclude by summarizing our main results.
An atom bouncing off a modulated mirror under the influence 
of gravity exhibits dynamical localization in position {\it and} momentum. 
However, this effect only occurs in a window of modulation. 
Our investigations show that bound systems may exhibit 
a rich dynamical behavior both in classical and quantum 
domain, which is substantially different
from the standard kicked rotator model. 
These conclusions motivate further theoretical and 
experimental studies of bound systems from the
view point of chaos.
Since our system in the absence of the driving force does not
contain any continuum of states it is cleaner than the microwave 
driven hydrogen atom.  
Currently available experimental technology 
allows us to observe this phenomenon. 

Our analysis of the atomic Fermi accelerator is based on a 
laser mirror for the atom and therefore makes use of the interaction
of the atom with an evanescent laser field. However, it is 
interesting to note that also magnetic mirrors~\cite{kn:hein}
for atoms exist and have produced many bounces. A modulation
of such a magnetic mirror could offer another possible 
realization of the atomic Fermi accelerator.  

We thank G. Alber, M. El Ghafar, R. Grimm, E. Mayr, 
P. T\"orm\"a, M. G. Raizen, V. Savichev 
and A. Zeiler for many fruitful discussions.
One of us FS thanks the Ministry of Science and Technology, Pakistan
for its continuous support. IBB is grateful to the Humboldt Foundation
and MF thanks the European Community for a grant in the framework
of the HCM network ``Non-Classical Light'' and INFM. 
This work was partially supported by DFG.

\end{multicols}

%\end{multicols}

\begin{references}
\bibitem{kn:tildas}We indicate the  physical variables with tildas and we will
                  drop all the tildas later when we introduce
                  the dimensionless variables.
\bibitem{kn:kbar}This value of $k^{\hspace{-2.1mm}-}$ is also consistent 
                 with the commutation relation $[z,p]=[\tilde z,\tilde p]\,\omega^3/(mg^2)
                 ={\it i}\hbar\omega^3/(mg^2)\equiv {\it i}k^{\hspace{-2.1mm}-}$
                 for the dimensionless canonical variables $z$ and $p$.
\end{references}
\end{document}